\documentclass[showpacs,aps,prd,twocolumn,nofootinbib]{revtex4}
\usepackage{amsmath}
\usepackage{epsfig}
\usepackage{subfigure}
\usepackage{float}
\usepackage{longtable}

\newcommand{\ep}{\varepsilon}

\newcommand{\vect}[1]{\vec{#1}}

\newcommand{\eqs}[1]{\begin{equation} \begin{split} #1\end{split} \end{equation} }

\newcommand{\ks}[1]{#1 \!\!\! \slash } 

\newcommand{\ga}{\gamma^5}

\newcommand{\ie}{{\it i.e.}}
\newcommand{\eg}{{\it e.g.}}
\newcommand{\etal}{{\it et al.}}

\newcommand{\ce}[1]{Eq.~(\ref{#1})}

\newcommand{\cf}[1]{{Fig.~\ref{#1}}}

\newcommand{\nn}{\nonumber}

\newcommand{\beq}[1]{
\begin{equation}\label{#1}}
\newcommand{\eeq}{\end{equation}}
\newcommand{\bea}[1]{
\begin{eqnarray}\label{#1}}
\newcommand{\eea}{\end{eqnarray}}

\newcommand{\eqsa}[1]{\begin{eqnarray}#1\end{eqnarray} }
\newcommand{\eq}[1]{\begin{equation}#1\end{equation} }
\newcommand{\eqsal}[1]{\begin{align}#1\end{align}}
\newcommand{\out}{\raise-3pt\hbox{\scriptsize    out}}

\begin{document}

\title{Production of a pion in association with a high-$Q^2$ dilepton  pair in $\bar p p $ 
annihilation
at GSI-FAIR}

\author
{J.P. Lansberg$^{a,b}$, B. Pire$^{b}$ and  L. Szymanowski$^{b,c,d}$
}

\preprint{CPHT--  hep-ph/yymmnnn}

\affiliation{
$^{a}$Institut f\"ur Theoretische Physik, Universit\"at Heidelberg, D-69120 Heidelberg, Germany\\
$^{b}$Centre de Physique Th\'eorique, \'Ecole Polytechnique, CNRS, 
91128 Palaiseau, France
\\$^{c}$Fundamental Interactions in Physics and Astrophysics, Universit\'e de Li\`ege, Belgium\\
$^{d}$Soltan Institute  for   Nuclear  Studies,  Warsaw,   Poland 
}

\begin{abstract}
We  evaluate the cross section for 
$\bar p p\to \ell^+\ell^- \pi^0$ in the forward direction and
for large lepton pair invariant mass.  In this kinematical region, the leading-twist amplitude
 factorises into a short-distance matrix element,  
 long-distance dominated antiproton Distribution Amplitudes  and proton to pion Transition Distribution 
Amplitudes (TDA). Using a modelling inspired from the chiral limit for these TDAs, we obtain a first 
estimate of this cross section, thus demonstrating that this process
 can  be measured  at GSI-FAIR.
\end{abstract}
\maketitle

%\section{Introduction}

Transition Distribution Amplitudes (TDAs)~\cite{TDA} are universal non-perturbative objects
describing the transitions between two different particles (~\eg~$p\to \pi$, $\pi\to\gamma$, 
$\pi\to \rho$). They appear  in the study of backward electroproduction of a
pion~\cite{Lansberg:2007ec}, of $\gamma^\star \gamma \to \rho \pi$ and
  $\gamma^\star \gamma \to \pi \pi$ reactions~\cite{TDApigamma-appl} 
 as well as in hard exclusive production of a $\gamma^\star \pi$ pair in $\bar p p $ annihilation:
\begin{equation}
\bar p (p_{\bar p })  p (p_p) \to \gamma^\star(q) \pi(p_\pi)\to \ell^+(p_{\ell^+})  \ell^-(p_{\ell^-})  \pi(p_\pi)
\label{process}
\end{equation}
at small $t=(p_\pi-p_p)^2$ (or at small $u= (p_\pi-p_{\bar p})^2$), which is the purpose of the present work.
 The TDAs are an extension of the concept of Generalised Parton Distributions (GPDs), as already
 advocated in~\cite{Frankfurt:1999fp}.
The proton to meson TDAs are defined from the Fourier transform  of a matrix element of a 
three-quark-light-cone operator between a proton and a meson state.
They obey QCD evolution equations
which follow from the renormalisation-group equation of  the
three-quark operator. Their $Q^2$ dependence is thus completely  under control.

Whereas in the pion to photon case, models used for 
GPDs~\cite{Tiburzi:2005nj,Broniowski:2007fs,Courtoy:2007vy,GPD_pion} 
could be applied to TDAs since they are defined from matrix elements of the same quark-antiquark operators, 
the situation is clearly different for the nucleon to meson  TDAs. Before estimates based on models such as the meson-cloud 
model~\cite{Pasquini:2006dv} become available, it is important to use as much model-independent information 
as possible. In~\cite{Lansberg:2007ec}, we derived constraints from the chiral limit
on the TDAs $p\to \pi$ and made a first evaluation of the cross section for the backward electroproduction
of a pion in the large-$\xi$ (or small $E_\pi$) region. Related processes 
were also recently studied in~\cite{Braun:2006td} similarly to what was proposed
in~\cite{Pobylitsa:2001cz}. In this work, we apply the same setting to evaluate the
cross sections for  $\bar p p\to \ell^+\ell^- \pi^0$
in the kinematical region accessible by GSI-FAIR~\cite{Spiller:2006gj}
 in the forward limit and at moderate energy of the meson.

In the scaling regime where $Q^2=q^2$ is of the order of $W^2 = (p_{\bar p} + p_{p})^2$, the amplitude for the process~(\ref{process})
at small $t$ -- or CM angle of the pion $ \theta^*_\pi$ 
close to 0 -- involves the  $p\to \pi$ TDAs $V^{p\pi}(x_{i}, \xi, \Delta^2)$, $A^{p\pi}(x_{i}, \xi, \Delta^2)$,
$T^{p\pi}(x_{i}, \xi, \Delta^2)$,
where $x_i$ ($i=1,2,3$) denote the light-cone-momentum fractions carried by participant 
quarks and $\xi$ is the skewedness parameter such that $2\xi =x_1+x_2+x_3$.
The amplitude is a convolution of the antiproton DAs, a 
perturbatively-calculable-hard-scattering amplitude and the $p\to \pi$ TDAs.
\begin{figure}[!ht]
\includegraphics[height=4.5cm]{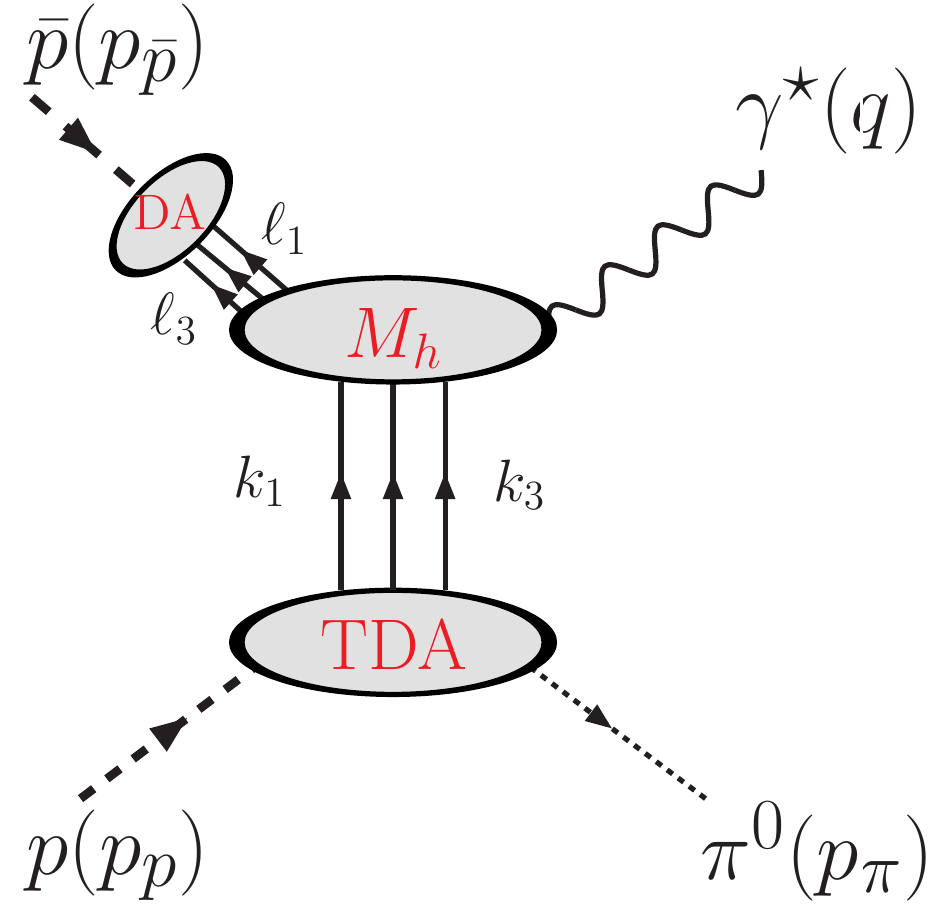}
\caption{The factorisation of the annihilation process $p\bar p \to \gamma^\star \pi$ into
antiproton-distribution amplitudes (DA), the hard-subprocess amplitude ($M_h$) and 
proton $\to$ pion transition distribution amplitudes (TDA) .
 }
\label{fig:fact-ampl}
\end{figure}

The momenta of the subprocess $\bar p p \to \gamma^\star \pi$ are defined as shown in~\cf{fig:fact-ampl}.
The $z$-axis is chosen along the colliding proton and antiproton 
and the $x-z$ plane is identified 
with the collision or hadronic plane. We define the 
light-cone vectors $p$ and $n$ 
such that $2~p.n=1$, as well as $P= (p_p+p_\pi)/2$, $\Delta=p_\pi -p_p$ and its 
transverse component $\Delta_T$ ($\Delta_T^2<0$).  $\xi$ is defined as $\xi=-\frac{\Delta.n}{2P.n}$.
We express the particle momenta  through a 
 Sudakov decomposition :
\eqsal{\label{eq:decomp_moment}
p_p=& (1+\xi) p+ \frac{M^2}{1+\xi}n\nn\\
p_{\bar p}=& \frac{2 M^2(1+\xi)}{\alpha}p+\frac{\alpha}{2 (1+\xi)}n\nn\\
p_\pi= &(1-\xi) p +\frac{m_\pi^2-\Delta_T^2}{1-\xi}n+ \Delta_T
}

\eqsal{\label{eq:decomp_momentb}
\Delta=& - 2 \xi p +\Big[\frac{m_\pi^2-\Delta_T^2}{1-\xi}- \frac{M^2}{1+\xi}\Big]
n + \Delta_T\nn\\
q\simeq & 2 \xi p +\frac{M^2}{W^2}(1+\xi)+ \Big[\frac{W^2}{1+\xi}-
\frac{m_\pi^2-\Delta_T^2}{1-\xi} \Big]n- \Delta_T\nn\\
\Delta_T^2=&\frac{1-\xi}{1+\xi} \Big( t-2 \xi \big[ \frac{M^2}{1+\xi}-\frac{m_\pi^2}{1-\xi}\big]\Big),
}
where $\alpha=W^2-2M^2+W\sqrt{W^2-4M^2}\simeq 2W^2$;
the approximate expression for $q$ is obtained with  $M\ll W$. 

For $\Delta_T=0$, $M\ll W$ and $m_\pi=0$, one gets 
\eqs{p_p&=(1+\xi)p, ~~~~~~ p_{\bar p}=\frac{W^2}{1+\xi} n, ~~~~~~ p_\pi=(1-\xi)p,\\
t&=\frac{2\xi M^2}{1+\xi}, ~~~~~~~~~~ \xi=\frac{Q^2}{2W^2-Q^2}.}
In the fixed-target mode, the maximal reachable value for $W^2 = 2M^2 +2 M E_{\bar p}$ at GSI will 
be $\simeq 30$ GeV$^2$ (for $E_{\bar p}=15$ GeV). The highest invariant mass 
of the photon could be $Q^2_{max}\simeq 30$ GeV$^2$. We refer to Ref~\cite{Adam}
for a complete discussion of the kinematically allowed domain. In terms of our notations, in the proton rest frame,
we have $p =\frac{M}{1+\xi} (1,0,0,-1)$ and $n =\frac{1+\xi}{4M} (1,0,0,1)$. Thus 
 $\xi \in [0.5, 1]$ corresponds to $|p^z_{\pi}|<M/3 \simeq 310 $ MeV in the laboratory frame at 
$\Delta_T=0$.

Let us now turn to the kinematics of ${\bar p} p \to \ell^+ \ell^- \pi^0$. In general, we
 have for the unpolarised differential cross section:
\eq{\label{eq:dsigma23}d\sigma= \frac{1}{2\sqrt{\lambda(W^2,M^2,M^2)} (2\pi)^5} |\overline{{\cal M}}|^2 d_3(PS).}
The 3-particle differential Lorentz invariant phase space (dLIPS), $d_3(PS)$,  can be decomposed
 into two 2-particle dLIPS (where $q$ is the momentum of the $\gamma^\star$):
\eqsal{
d_3(PS)=&\delta^4(p_p+p_{\bar p}-p_{\ell^+}-p_{\ell^-}-p_\pi) \frac{d^3\vect p_{\ell^+}}{2 p^0_{\ell^+}}
\frac{d^3\vect p_{\ell^-}}{2 p^0_{\ell^-}} \frac{d^3\vect p_\pi}{2 p^0_\pi}\nn\\\nn
=&\delta^4(p_p+p_{\bar p}-q-p_\pi)\frac{d^3\vect p_\pi}{2 p^0_\pi} \frac{d^3\vect q}{2 q^0} \times dQ^2 
\\
\times&\delta^4(q-p_{\ell^+}-p_{\ell^-}) \frac{d^3\vect p_{\ell^+}}{2 p^0_{\ell^+}}
\frac{d^3\vect p_{\ell^-}}{2 p^0_{\ell^-}}.}
In the $\bar p p $ CM, we have:
\eq{\delta^4(p_p+p_{\bar p}-q-p_\pi) \frac{d^3\vect p_\pi}{2 p^0_\pi} \frac{d^3\vect q}{2 q^0}
=\frac{d\Omega^\star_\pi}{8 W^2} \sqrt{\lambda(W^2,Q^2,m^2_\pi)} }
and in the $\ell^+\ell^-$ CM , we have ($m_\ell\simeq 0$):
\eq{\delta^4(q-p_{\ell^+}-p_{\ell^-}) \frac{d^3\vect p_{\ell^+}}{2 p^0_{\ell^+}}
\frac{d^3\vect p_{\ell^-}}{2 p^0_{\ell^-}}
= \frac{d\Omega_\ell}{8}=\frac{d\cos\theta_\ell d\varphi_\ell}{8}.}
Expressing  $t=(p_\pi-p_p)^2$ in terms of $\cos \theta^\star_\pi$~\cite{Borodulin:1995xd}, we get
\eq{
dt=  
%M^2+m_\pi^2+\frac{s+Q^2-m^2_\pi}{2}+
\frac{d\cos \theta^\star_\pi}{2 W^2}\sqrt{\lambda(W^2,M^2,M^2)} 
\sqrt{\lambda(W^2,Q^2,m^2_\pi)}}
Altogether, by integrating on $\varphi^\star_\pi$ and on $\varphi_\ell$ , 
\eq{\label{eq:dsigmadtdcosth}
  \frac{d \sigma}{dtdQ^2 d\cos\theta_\ell}=\frac{\int d\varphi_\ell |\overline{{\cal M}^{\bar p p \to \ell^+ \ell^- \pi^0}}|^2}{64 W^2 (W^2-4M^2) (2\pi)^4}
}
to be compared with the cross section for $\bar p p \to \gamma^\star \pi^0$
\eqs{\label{eq:dsigmadt}
\frac{d \sigma}{dt}=& \frac{|\overline{{\cal  M}^{\bar p  p\to\gamma^\star \pi^0}}|^2}{16 \pi W^2 (W^2 -4M^2)} .
}

At $\Delta_T=0$, the leading-twist TDAs for the $p \to \pi^0$ transition, $ V^{p\pi^0}_{i}\!\!(x_i,\xi, \Delta^2)$, 
$A^{p\pi^0}_{i}\!\!(x_i,\xi, \Delta^2)$ and 
$T^{p\pi^0}_{i}\!\!(x_i,\xi, \Delta^2)$  are defined  as
(see Appendix for details) :
\eqsal{\label{eq:TDApi0proton}
  {\cal F}\Big(\langle     \pi^0(p_\pi)|\, &\epsilon^{ijk}u^{i}_{\alpha}(z_1 n) 
u^{j}_{\beta}(z_2 n)d^{k}_{\gamma}(z_3 n)
\,|P(p_p,s_p) \rangle \Big)=   
\nn\\
\frac{i}{4}\frac{f_N}{f_\pi}\Big[ &V^{p\pi^0}_{1} (\ks p C)_{\alpha\beta}(u^+(p_p,s_p))_{\gamma}
\\& +A^{p\pi^0}_{1} (\ks p\gamma^5 C)_{\alpha\beta}(\gamma^5 u^+(p_p,s_p))_{\gamma} 
\nn\\& +T^{p\pi^0}_{1} (\sigma_{p\mu} C)_{\alpha\beta}(\gamma^\mu u^+(p_p,s_p))_{\gamma}\Big]\;,\nn}
where  $\sigma^{\mu\nu}= 1/2[\gamma^\mu, \gamma^\nu]$, $C$ is the charge 
conjugation matrix, $f_\pi = 131$ MeV is the pion decay constant  and $f_N \sim 5.2\cdot 10^{-3}$ GeV$^2$. 
 $u^+$ is the large component of the nucleon spinor :
$u(p_p,s_p)=(\ks n \ks p + \ks p \ks n) u(p_p,s_p) = u^-(p_p,s_p)+u^+(p_p,s_p)$
with $u^+(p_p,s_p)\sim \sqrt{p_p^+}$ and $u^-(p_p,s_p)\sim \sqrt{1/p_p^+}$.

For the three TDAs $V^{p\pi^0}$, $A^{p\pi^0}$ and $T^{p\pi^0}$, contributing in the 
limit $\Delta_T \to 0$, we use the following expressions for  $\Delta_T=0$ and large $\xi$
(see Appendix):
\eqsal{\label{eq:softp}
\{V^{p\pi^0}_1,A^{p\pi^0}_1,T^{p\pi^0}_1\}&(x_1,x_2,x_3,\xi,\Delta^2) = \\
 &\frac{1}{4 \xi}  \{V^p, A^p,3 T^p\} (\frac{x_1}{2\xi},\frac{x_2}{2\xi},\frac{x_3}{2\xi}),\nn}
where $V^{p}$, $A^{p}$ and $T^{p}$ are the proton DAs~\cite{CZ}.

At the leading order in $\alpha_s$ and at $\Delta_T=0$, the amplitude 
${\cal M}_\lambda^{s_ps_{\bar p}}$  for 
$\bar p(p_{\bar p},s_{\bar p}) p(p_{p},s_p) \to \gamma^\star(q,\lambda) \pi^0(p_\pi)$ reads
\eq{\label{eq:ampl-bEPM1}
{\cal M}_\lambda^{s_p s_{\bar p}}=
-i 
\frac{(4 \pi \alpha_s)^2 \sqrt{4 \pi \alpha_{em}} f_{N}^2}{ 54 f_{\pi}Q^4} {\cal S}_\lambda^{s_ps_{\bar p}}{\cal I}}with ${\cal S}_\lambda^{s_ps_{\bar p}}=\bar v^+(p_{\bar p},s_{\bar p}) \ks \ep^\star(\lambda) \gamma^5 u^+(p_p,s_p)$
and
\eqs{\!\!\!{\cal I}\!=\!\!\!\!\!\int\limits^{1+\xi}_{-1+\xi} \! \! \! [dx] \int\limits_0^1 \! \![dy]
\Bigg(2\sum\limits_{\alpha=1}^{7} R_{\alpha}+
\sum\limits_{\alpha=8}^{14} R_{\alpha}\Bigg),}
where $[dx]=dx_1 dx_2 dx_3\delta(2\xi - \sum_k x_k)$ and $[dy]=dx_1 dx_2 dx_3\delta(1 - \sum_k y_k)$; 
the coefficients $R_{\alpha}\,(\alpha=1,...,14)$ exactly correspond to $T_{\alpha}$
in~\cite{Lansberg:2007ec} after the replacement $-i\epsilon \to i \epsilon $ due to the presence 
of the $\gamma^\star$ in the final  instead of initial state.
Even though the TDA formalism can be applied at any value of $\xi$ (or $E_\pi$),
 we have for now at our disposal estimates for the $p\to \pi$ 
TDAs only at large $\xi$. In the following, we 
shall therefore limit ourselves to the computation of the cross section 
for reaction (\ref{process}) in this region.
At large $\xi$, the ERBL regime ($x_i>0$)  covers most of the integration domain.  
Therefore  it is legitimate to approximate the cross section only from the ERBL 
contribution, \ie~when the integration range of 
the momentum fractions is restricted to $[0,2\xi]$.

The differential cross section for  unpolarised protons and antiprotons is 
calculated as usual using \ce{eq:dsigmadt}. from the 
averaged-squared amplitudes, 
\eqs{
|\overline{{\cal  M}_{\lambda \lambda'}}|^2= \frac{1}{4}
\sum_{s_p s_{\bar p}} {\cal M}^\lambda_{s_ps_{\bar p}}({\cal M}^{\lambda'}_{s_ps_{\bar p}})^*.
}
$|\overline{{\cal  M}_{00}}|^2$ vanishes at the leading-twist accuracy, 
as in the nucleon-form-factor case. The same is true for $|\overline{{\cal  M}_{+-}}|^2$ and
$|\overline{{\cal  M}_{0+}}|^2$, etc., since the  $x$ and $y$ directions are not distinguishable
when  $\Delta^2_T$ is vanishing. We then define $|\overline{{\cal  M}_T}|^2\equiv|\overline{{\cal  M}_{++}}|^2+|\overline{{\cal  M}_{--}}|^2$.

To compute $\cal I$, we need to choose models for the DAs and the deduced TDAs.
For the sake of coherence with experimental data, we shall use  
reasonable parametrisations of CZ~\cite{CZ} 
 and KS~\cite{KS}, which are both based on an analysis of QCD sum rules. 
For CZ, they are
\eqsa{\label{eq:DACZ}
V^p(x_i)&=&\varphi_{as} [11.35 (x_1^2+x_2^2)+8.82 x_3^2-1.68 x_3 -2.94],\nn\\
A^p(x_i)&=&\varphi_{as} [6.72 (x_2^2-x_1^2)],\\
T^p(x_i)&=&\varphi_{as} [13.44 (x_1^2+x_2^2)+4.62 x_3^2+0.84 x_3 -3.78],\nn
}
and for KS (which we used in \cf{fig:cross-section})
\eqsa{\label{eq:DAKS}
V^p(x_i)&=&\varphi_{as} [17.64 (x_1^2+x_2^2)+22.68 x_3^2-6.72 x_3 -5.04],\nn\\
A^p(x_i)&=&\varphi_{as} [2.52 (x_2^2-x_1^2) + 1.68(x_2-x_1)],\\
T^p(x_i)&=&\varphi_{as} [21.42 (x_1^2+x_2^2)+15.12 x_3^2+0.84 x_3 -7.56],\nn
}
and we evaluate our model TDAs from \ce{eq:softp}. This gives
${\cal  I}\simeq 1.28 \cdot 10^{5} $ for CZ  and ${\cal  I}\simeq 2.15\cdot 10^{5}$ 
for  KS; this yields an induced uncertainty of order 3 for our estimates of the cross section. 
In the following, we use $\alpha_s =0.3 $ as suggested in~\cite{CZ}.

Altogether, we have the following analytic results for the dominant ERBL contribution:
\eqs{
|\overline{{\cal  M}_T}|^2 = \frac{(4 \pi \alpha_s)^4 (4 \pi \alpha_{em}) f_{N}^4}{ 54^2 f^2_{\pi}}
\frac{2(1+\xi)|{\cal  I}|^2}{\xi Q^6}\; . 
}
 From this, we
straightforwardly obtain $\frac{d \sigma}{dt}$, whose $W^2$ evolution is displayed 
on \cf{fig:cross-section} (a) at $\Delta_{T}=0$  for the two extreme values of meson 
longitudinal momentum  where one may trust the soft pion limit,
corresponding to $p_{\pi}^z = 0 $ or $|p^z_{\pi}| = M/3 $ in the laboratory frame
($\xi = 1$ or $ 1/2$).

For the process (\ref{process}), the averaged-squared
amplitude is:
\eqsal{
|\overline{{\cal  M}^{\bar p p \to \ell^+ \ell^- \pi^0}}|^2= \frac{1}{4}
\sum_{s_p,s_{\bar p},\lambda,\lambda'} {\cal M}^\lambda_{s_ps_{\bar p}}\frac{1}{Q^2}{
\cal L}^{\lambda\lambda'}
\frac{1}{Q^2}
({\cal M}^{\lambda'}_{s_ps_{\bar p}})^*,\nn}
with ${\cal L}^{\lambda\lambda'}=e^2{\rm Tr} (\ks p_{\ell^-} \ks \ep(\lambda) \ks p_{\ell^+} \ks\ep^\star(\lambda'))$. Integrating on the lepton azimuthal angle $\varphi_{\ell}$, we have
\eqsal{
\int d\varphi_\ell |\overline{{\cal  M}^{\bar p p \to \ell^+ \ell^- \pi^0}}|^2=
|\overline{{\cal  M}_T}|^2 \frac{2\pi  e^2 (1+\cos^2 \theta_{\ell})}{Q^2},}
from which we get, via \ce{eq:dsigmadtdcosth} and integrating over $\theta_\ell$
the differential cross section displayed in \cf{fig:cross-section} (b).
\begin{figure}[t!]
\includegraphics[scale=.7,clip=false]{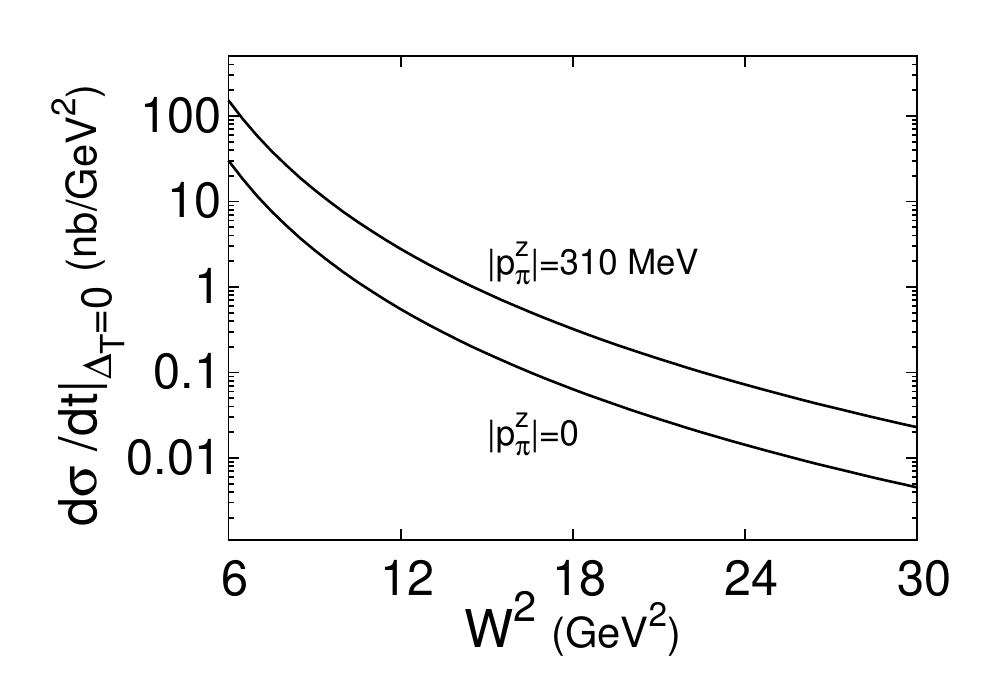}
\includegraphics[scale=.7,clip=false]{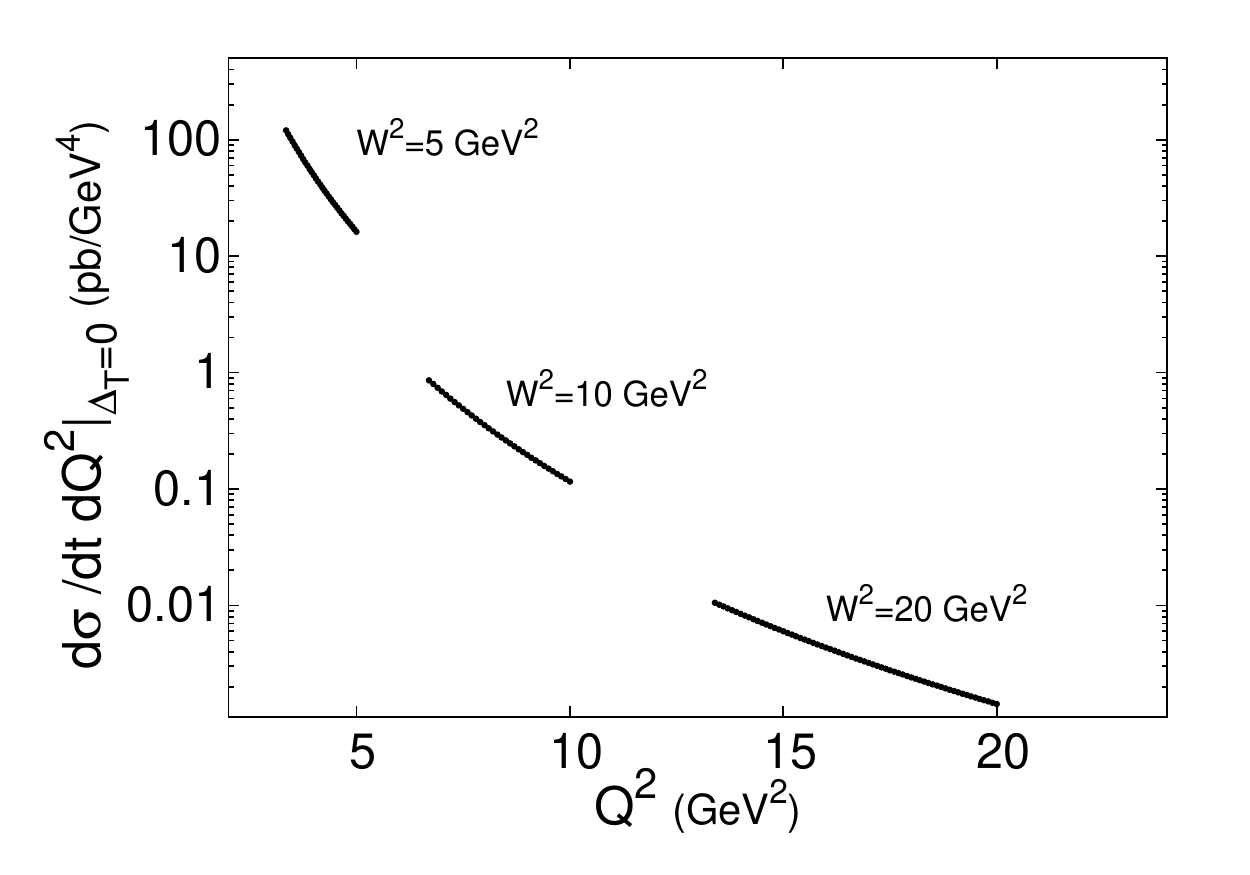}
\caption{
(a) Differential cross section $d\sigma/dt$ for $\bar p p \to \gamma^\star \pi^0$ 
as a function a $W^2$ for $|p^z_{\pi}| = 0$ (lower curve) and
$|p^z_{\pi}| =M/3 $.
(b) Differential cross section $d\sigma/(dtdQ^2)$ for $\bar p p \to \ell^+ \ell^- \pi^0$ 
as a function of $Q^2$  for various beam energies.}
\label{fig:cross-section}
\end{figure}

Although the cross sections are evaluated at $\Delta_T=0$, we do not 
anticipate any dramatic
$\Delta_T$-dependence of the TDAs below a few hundred MeV, so that 
our estimates are likely 
to be valid in a not-too-narrow  $\Delta_T$ region.  
To evaluate a magnitude of the integrated cross section we take as an example
the kinematical region with $W^2=10$ GeV$^2$ in Fig.2b accompanied by the $Q^2$
window $7$~GeV$^2$ $ < Q^2 <8$~GeV$^2$, which corresponds to $\xi \approx 1/2$
 or to the pion momentum of the order 310MeV.
Integrating over this $Q^2$-bin 
 and in a $t-$bin corresponding to $\Delta_T< 500$ MeV leads then to a cross section around 100 femtobarns. 
Such a cross section 
 is sizable and seems to be accessible to experimental setups such as PANDA
with the designed value of the FAIR luminosity.

The calculations done till now and which involve the proton $\to \pi^0$ TDA 
are valid for the small $t$ region.
Let us however stress that - due to the charge symmetry - an identical 
result will be obtained in the small $u$ region but with the
 $\bar p \to \pi^0$ TDA. 
In the laboratory frame at GSI-FAIR, this second region is quite different from 
the previous one since the $\pi^0$ meson is boosted in the forward direction.
A precise detection of the particles of the final state, either in the proton or the antiproton "fragmentation" kinematics, will depend on the detector performances in the respective regions.

In conclusion, we have demonstrated that the study of proton-antiproton exclusive annihilation
into a lepton pair and a pion is  feasible at large values of the lepton pair invariant mass
in the forthcoming PANDA or PAX experiments at GSI-FAIR. We believe that
such a study will bring unique information
about the inner structure of the proton, and particularly about the pion content of a proton,
provided that the predictions (scaling behaviour, angular dependence of the lepton pair) of  the factorised framework
used here are shown to be valid. Expressing  the cross section of this process  as the
convolution of a hard-scattering amplitude for quarks with a photon with DAs and 
 proton  to pion TDAs will allow to evaluate these new hadronic matrix elements which contain 
 much information about the confinement dynamics. It has to be emphasised that the same
hadronic matrix elements appear also in a similar description of backward electroproduction
of a pion~\cite{Lansberg:2007ec}.

Note that other channels are also of much interest, such as $\bar p p  \to l^+ l^- \eta$ or
 $\bar p p \to l^+ l^- \rho^0$ . The theoretical  treatment of the $\eta$ case is identical 
 to the one for the $\pi^0$  case, but for the isosinglet nature of $\eta$.
 In the $\rho^0$ case, one should distinguish between the  longitudinally polarised meson where 
 the TDA has the same structure  as for the $\pi^0$ and the transversally polarised case which leads
  to  more TDAs. Needless to say, we are strongly lacking of model estimates  for these $p\to \eta$ 
  and $p\to \rho$ TDAs but their experimental  determination (or at least the measurement of their
   ratios to the  $p \to \pi$ TDAs ) opens a fascinating window on the properties of the  sea quarks
    in the proton wave function.

We are thankful to V. Braun, T. Hennino and F. Maas  for useful and stimulating discussions.
This works is  supported  by  the Polish grant
1~P03B~028~28, the French MAEE  Eco-Net program, the EU contract RII3-CT-2004-506078
and the FNRS (Belgium).

{\bf Appendix.} We now  derive the general limit
of the three contributing TDAs at $\Delta_T=0$ in the soft-pion limit, when $\xi$ gets close
to 1. In that limit, the soft-meson theorem~\cite{AD} 
derived from current algebra
applies~\cite{Pobylitsa:2001cz}, which allows us to express these 3 TDAs in terms  of the 3 Distribution
Amplitudes (DAs) of the corresponding baryon. 
Conventionally~\cite{CZ}, the three proton  DAs are defined through the decomposition of the 
following matrix element of the 3-quark operator 
 in terms of three invariant functions of the scalar product of
the light-like separation $z_i n\equiv\tilde z_i$ with the proton 
momentum $p_p$, $V^p(\tilde z_i.p_p)$,$A^p(\tilde z_i.p_p)$ and $T^p(\tilde z_i.p_p)$,  
\eqs{&\langle 0|u_\alpha(z_1 n)u_\beta(z_2 n)d_\gamma(z_3 n)|p_p)\rangle=\frac{1}{4}f_N \times \\
 \Big[
&V^p(\tilde z_i.p_p) ( \ks p_p C)_{\alpha \beta} (\gamma^5 u^+(p_p,s_p))_\gamma 
\\ +&A^p(\tilde z_i.p_p) (\ks p_p \gamma^5 C)_{\alpha \beta} u^+(p_p,s_p)_\gamma
\\ +&T^p(\tilde z_i.p_p) (\sigma_{p_p  \mu}\,C)_{\alpha \beta} 
(\gamma^\mu \gamma^5 u^+(p_p,s_p))_\gamma
\Big] \;.
\label{eq:DA}}
The latter functions satisfy $V^p(\tilde z_i.p_p=0)=T^p(\tilde z_i.p_p=0)=1$ and  $
A^p(\tilde z_i.p_p=0)=0$,
which provides the interpretation of $f_N$ as the value of the proton wave function at the origin.
To go to  momentum space one writes a  Fourier transform~\cite{KS} 
which  enables to define functions of momentum fractions $x_i$ ($F=V^p,A^p,T^p$) 
($[d \tilde z.\cdot]=d(\tilde z_1.\cdot) d(\tilde z_2.\cdot) d(\tilde z_3.\cdot)$)
:
\eqs{\label{eq:fourier_DA}
\tilde F(x_i)  \equiv \int^{\infty}_{-\infty} 
\frac{[d\tilde z.p_p]}{(2\pi)^3} e^{i\Sigma_k x_k \tilde z_k.p_p }F(\tilde z_i.p_p).}
Inspired by~\cite{Pobylitsa:2001cz}, which considered the related case of the 
distribution amplitude of the proton-meson system, we use the soft pion 
theorems~\cite{AD} to write:
\eqs{
&\langle \pi^a(p_\pi)  |{\cal O}| P(p_1,s_1)\rangle = -\frac{i}{f_\pi} \langle 0  | [ Q^a_5,  {\cal O}]  | P(p_1,s_1) \rangle,% \\ \nonumber
\label{eq:soft-theorem}
}
where we neglected the nucleon pole term, which does not contribute at threshold.

For the transition $p\to \pi^0$, $Q^a_5=Q^3_5$ and ${\cal O}=u_\alpha u_\beta d_\gamma$. 
Since the commutator of the chiral charge $Q_{5}$ with the quark field $\psi$ ($\tau^a $ being the Pauli 
matrix)
\begin{equation}
[Q_{5}^a, \psi] = - \frac{\tau^a}{2} \gamma^5 \psi\;,
\label{eq:chiral-trans}
\end{equation}
the first term in the rhs of \ce{eq:soft-theorem} gives three terms from 
$(\ga u)_\alpha u_\beta d_\gamma$, $u_\alpha (\ga u)_\beta d_\gamma$ and $u_\alpha u_\beta (\ga d)_\gamma$.
The corresponding multiplication by $\ga$ (or $(\ga)^T$ when it acts on the index $\beta$) on the 
vector and axial-vector structures of the DA (\ce{eq:DA}) gives two terms which cancel and the third one,
which remains, is the same as the one for the TDA up to the modification that on the DA decomposition,
$p_p$ is the proton momentum, whereas for the TDA one, $p$ is the light-cone projection of $P$, $\ie$ half
the proton momentum if one neglects the pion one. This introduces a factor $2\xi$ in the relations
between the 2 DAs $A^p$ and $V^p$ and the 2 TDAs $V^{p\pi^0}_{1}$ and $A^{p\pi^0}_{1}$.
%, which is partially canceled though by the factor one half in \ce{eq:chiral-trans}. 

To what concerns the tensorial structure multiplying $T^p$,  the three 
terms are identical at leading-twist accuracy and correspond to the structure multiplying
$T^{p\pi^0}_{1}$, this gives a factor 3. We eventually have the soft limit
for our three TDAs at $\Delta_T=0$:
\eqs{\label{eq:softp_coordinate}
(V^{p\pi^0}_1,A^{p\pi^0}_1,T^{p\pi^0}_1)(\tilde z_i.p) =\xi (V^p,A^p,3 T^p) (\tilde z_i.p_p)}

We will derive now this relation in the momentum representation of DAs and TDAs.
To do so, we start from translational invariance which  implies
\eqs{
\langle 0|u_\alpha(\tilde z_1+a)u_\beta(\tilde z_2+a)d_\gamma(\tilde z_3+a)|p_p\rangle=\\e^{-i a.p_p}
\langle 0|u_\alpha(\tilde z_1)u_\beta(\tilde z_2)d_\gamma(\tilde z_3)|p_p\rangle,}
and thus $F((\tilde z_i+a).p_p)= e^{-i a.p_p}F(\tilde z_i.p_p)$. In momentum space, we correspondingly get
\eqsa{\label{eq:trans_inv}
\tilde F(x_i)&=&
\int^{\infty}_{-\infty} 
\frac{[d(\tilde z+a).p_p]}{(2\pi)^3} e^{i\Sigma_k x_k(\tilde z_k+a).p_p }F((\tilde z_i+a).p_p)\nn
\\&=&e^{i(\Sigma_k x_k -1) a.p_p} \tilde F(x_i).
}
This condition is conveniently expressed by the following redefinition:
$\tilde F(x_i)= \delta(\sum_k x_k -1)F(x_i)$. The inverse Fourier transform is then written as ($[dx]=dx_1dx_2dx_3$):
\eqsa{
F(\tilde z_i.p_p)=\int_0^1 [dx]\; e^{-i \Sigma_k x_k\tilde z_k.p_p} \delta(\sum_k x_k -1)F(x_i).}
The normalisation conditions then reads
\eqs{
\int_0^1 [dx]\; \delta(\sum_k x_k -1)(V^p,A^p,T^p)(x_i)=(1,0,1).}

Note that the delta function  insuring translational invariance is exactly the
one expected from the interpretation that $x_i$ be the momentum fraction carried
 by the quark $i$ off the proton of momentum $p_p$. 
This shows that
the natural conjugate variable to the momentum fractions  $x_i$'s are the $\tilde z_i.p_p$'s, 
\ie~the spatial separation dotted by the proton momentum. 
Indeed,  $p_p$ enters in the exponential of the rhs of~\ce{eq:trans_inv},
via the conjugate variable to $x_i$, $\tilde z_i.p_p$, and as the initial-state momentum of
 the matrix element. Another choice than $\tilde z_i.p_p$ would not have provided the correct support for the $x_i$'s.

The case of TDAs
is similar except for the choice of the natural conjugate variable. We start with  translational invariance:
\eqs{
\langle B(p_B)|\epsilon^{ijk}u^i_\alpha(\tilde z_1+a)u^j_\beta(\tilde z_2+a)d^k_\gamma(\tilde z_3+a)|A(p_A)
\rangle=\\e^{-i a.(p_A-p_B)}
\langle B(p_B)|\epsilon^{ijk}u^i_\alpha(\tilde z_1)u^j_\beta(\tilde z_2)d^k_\gamma(\tilde z_3)|
A(p_A)\rangle.\nn}
Now let us define a Fourier transform without specifying the momentum $p_0$ to which we dot 
the spatial separation:
\eqsal{\label{eq:trans_inv_tda}
&\tilde F_{A\to B}(x_i)=
\int^{\infty}_{-\infty} \!\!\!
\frac{[d((\tilde z+a).p_0)]}{(2\pi)^3} e^{i\Sigma_k x_k(\tilde z_k+a).p_0 }\times \\
&F_{A\to B}((\tilde z_i+a).p_0)%\\
=e^{i(\Sigma_k x_k a.p_0-a.(p_A-p_B))} 
 \tilde F_{A\to B}(x_i).\nn
}
The condition derived from translational invariance 
would then be satisfied by $\delta(\sum_k x_k a.p_0-a.(p_A-p_B))$. In order
to get the correct support, \ie~$\delta(2\xi - \sum_k x_k)$, for a translation $a$ along $n$, 
we have to choose $p_0$ such that
$2 \xi=\frac{ n.(p_A-p_B)}{n.p_0}=-\frac{ n.\Delta}{n.p_0}$,
which is satisfied  by $p_0=(p_A+p_B)/2=P$ or a light-cone vector $p$ ($p^2=0$) such that $P.n=p.n$.
We choose $p$ and have 
\eqsal{&
\delta(2\xi - \sum_k x_k) F_{A\to B}(x_i)\equiv{\cal F}\big(F(z_i (n.p))\big) \equiv \\&
(n.p)^3\int^{\infty}_{-\infty} 
\frac{[dz]}{(2\pi)^3} e^{i\Sigma_k x_k z_k (n.p) }F(z_i (n.p)).\nn}

Let us now use the Fourier transform~\ce{eq:fourier_DA} on both side of~\ce{eq:softp_coordinate}. 
Defining $\alpha$ such that $p_p.n=\alpha P.n=\alpha p.n$, we get for instance for the $V$'s:
\eqsal{ &\delta(\sum_k x_k -1) V^p(x_i)=
\int^{\infty}_{-\infty} 
\frac{[d\tilde z.p_p]}{(2\pi)^3} e^{i\Sigma_k x_k \tilde z_k.p_p }
V^{p\pi^0}_1(\tilde z_i.p)\nn\\
&=\alpha^3 (n.p)^3 \int^{\infty}_{-\infty} 
\frac{[dz]}{(2\pi)^3} e^{i\Sigma_k x_k z_k \alpha (n.p) } V^{p\pi^0}_1(z_i(n.p))\nn\\
&= \alpha^3  \delta(2\xi - \sum_k (\alpha x_k)) V^{p\pi^0}_1(\alpha x_i).}
We conclude that translational invariance imposes naturally, through the delta function, 
that $\alpha=2\xi$; the change of variable $x'_i=\alpha x_i$ then yields
\eqsal{\label{eq:softpbis}
\{V^{p\pi^0}_1,A^{p\pi^0}_1,T^{p\pi^0}_1\}&(x_1,x_2,x_3,\xi,\Delta^2) = \\
 &\frac{1}{4 \xi}  \{V^p, A^p,3 T^p\} (\frac{x_1}{2\xi},\frac{x_2}{2\xi},\frac{x_3}{2\xi}),\nn}

%%%%%%%%%%%%%%%%%%%%%%%%%%%%%%%%%%%%%%%%%%%%%%%%%%%%%%%%%%%%%%%%%%%%%%%%%%%%

\end{document}